\documentclass[twocolumn,nofootinbib,amsmath,amssymb]{revtex4}
\usepackage[colorlinks,
            linkcolor=blue,
            anchorcolor=blue,
            citecolor=blue]{hyperref}
\usepackage{natbib}
\usepackage{doi}
\usepackage{url}
\usepackage{exscale}
\usepackage{amsmath}
\usepackage{booktabs}
\usepackage{graphicx}
\usepackage{exscale}
\usepackage{relsize}
\usepackage{graphicx}
\usepackage{dcolumn}
\usepackage{bm}
\usepackage{multirow}
\usepackage{CJK}
\usepackage{diagbox}
\usepackage{subfigure}
\usepackage{slashed}
\usepackage{makecell}
\usepackage{adjustbox}
\usepackage{longtable}
\usepackage{multirow}
\usepackage{amsmath}
\usepackage{subfloat}

\makeatletter

\newcommand{\Rmnum}[1]{\expandafter\@slowromancap\romannumeral #1@}
\def \nl {\nonumber\\}
\makeatother

\begin{document}

\title{Nature of $X(3872)$ from its radiative decay}
\author{Shuo-Ying Yu $^{1}$}
\author{Xian-Wei Kang $^{1, 2}$}\email{xwkang@bnu.edu.cn}

\affiliation{$^1$ Key Laboratory of Beam Technology of the Ministry of Education,
College of Nuclear Science and Technology, Beijing Normal University, Beijing 100875, China\\
$^2$ Institute of Radiation Technology, Beijing Academy
of Science and Technology, Beijing 100875, China}

\begin{abstract}
We study the radiative decay of $X(3872)$ based on the assumption that $X(3872)$ is regarded as a $c\overline{c}$ charmonium with quantum number $J^{PC}=1^{++}$ ($J,\,P,\,C$ represent the spin, parity and charge conjugation, respectively). The form factors of $X(3872)$ transitions to $J/\psi\gamma$ and $\psi'\gamma$ ($\psi'$ denotes $\psi(2S)$ throughout the paper) are calculated in the framework of the covariant light-front quark model. The phenomenological wave function of a meson depends on the parameter $\beta$, whose inverse essentially describes the confinement scale. In the present work, the parameters $\beta$ for the vector $J/\psi$ and $\psi'$ mesons will be determined through their decay constants, which are obtained from the experimental values of their partial decay widths to the electron-positron pair. For $X(3872)$, we determined the value of $\beta$ by the decay width of $X(3872)\rightarrow \psi'\gamma$. Then, we examined the width of $X(3872)\to J/\psi\gamma$ in a manner of parameter-free prediction and compared it with the experimental value. As a result, an inconsistency or contradiction occurs between the widths of $X(3872)\to J/\psi\gamma$ and $X(3872)\to \psi'\gamma$. We thus conclude that $X(3872)$ cannot be a pure $c\overline c$ resonance and that other components are necessary in its wave function.
\end{abstract}

\maketitle
\section{introduction}
Traditionally there exists two types of hadrons in nature: mesons composed of quarks and anti-quarks and baryons composed of three quarks. As a key topic in hadron physics, researchers have put much effort into both theoretical studies and experimental searches for exotic states beyond the aforementioned configurations, such as tetraquark, pentaquark, glueball and hybrid states. In addition, some peaks are only a manifestation of the threshold effects \cite{Haidenbauer:2015yka,Guo:2019twa}. In 2003, the Belle Collaboration observed a narrow resonance state $X(3872)$ in the mass spectrum of $\pi^+\pi^-J/\psi$ \cite{Belle:2003nnu}, which is considered to be the first candidate of an exotic state. Later, many measurements were made on its mass and decay properties. The LHCb Collaboration determined the quantum number of $X(3872)$ as $J^{PC}=1^{++}$ (with $J,\,P,\,C$ denoting the spin, parity and charge conjugation value, respectively) by performing a full five-dimensional amplitude analysis of the angle correlation \cite{LHCb:2013kgk,LHCb:2015jfc}. The signals of $X(3872)\rightarrow D^0\bar{D}^{*0}$ \cite{BESIII:2020nbj}, $\pi^+\pi^-J/\psi$ \cite{LHCb:2020fvo}, $\omega J/\psi$ \cite{BESIII:2019qvy}, $\pi^0\chi_{c1}$ \cite{BESIII:2019esk}, $J/\psi\gamma$ \cite{Belle:2011wdj}and $\psi'\gamma $\cite{LHCb:2014jvf} were also observed. For more details, the Particle Data Group (PDG) can be referenced \cite{ParticleDataGroup:2022pth}. We will focus on the radiative decays $X(3872)\to J/\psi\gamma$ and $\psi'\gamma$. The most recent measurement of their ratio $\frac{\text{Br}(X(3872)\rightarrow \psi(2S)\gamma)}{\text{Br}(X(3872)\rightarrow J/\psi\gamma)}$ (Br denotes the branching ratio) is from the BESIII collaboration, with a value $<0.59$ at the 90\% confidence level (C.L.) \cite{BESIII:2020nbj}. It agrees with the upper limit $<2.1$ at 90\% C.L. set by the Belle collaboration while marginally agreeing with the BaBar value $3.4 \pm 1.4$ \cite{BaBar:2008flx} and challenging the LHCb value $2.46 \pm 0.70$ \cite{LHCb:2014jvf} within two standard deviations. Taking into account the model predictions, BESIII disfavors the interpretation of a pure charmonium state compared to other interpretations \cite{BESIII:2020nbj}. Until 2020, significant progress was made in measuring the width of $X(3872)$ by LHCb \cite{LHCb:2020xds,LHCb:2020fvo}. Now, the PDG average is $\Gamma_{X(3872)}=(1.1 \pm 0.21)$ MeV.

As shown in Ref.~\cite{Kang:2016jxw}, the data on the line shape of $X(3872)$ are not sufficient to determine its pole structure from the perspective of compositeness \cite{Li:2021cue,Baru:2003qq,Kinugawa:2023fbf}: either a bound state or a virtual state of $\bar D D^*$ is possible for $X(3872)$.
See also the review \cite{Guo:2017jvc} for more discussions. Consequently, it is necessary to exploit the $X(3872)$ decay information to investigate its inner structure from another point of view. Among the various decay channels, radiative decay can be measured experimentally and treated in theory in an easier way. In this work, we assume that $X(3872)$ is a pure charmonium state. The plausibility of such a hypothesis is tested by comparing the calculated width $\Gamma(X(3872)\rightarrow J/\psi(\psi')\gamma)$ to the experimental widths. The form factors describing the radiative transition of $X(3872)\to J/\psi, \psi'$ are calculated in the covariant light-front quark model (CLFQM). This quark model involves a free parameter $\beta$ that appears in the wave function of a hadron. The values of $\beta$ for $J/\psi$ and $\psi'$ will be fixed by their decay constants, or more precisely speaking, by the values of $\Gamma(J/\psi (\psi')\to e^+e^-)$. Taking the decay width $\Gamma(X(3872)\rightarrow \psi'\gamma)$ \cite{Workman:2022ynf} as input, we can fix the parameter $\beta$ for $X(3872)$. Then, the calculation of the width $\Gamma(X(3872)\rightarrow J/\psi\gamma)$ will be a parameter-free prediction. The comparison between it and the experimentally measured width will definitely provide a criterion for how the interpretation of a conventional charmonium works for $X(3872)$.

This paper is organized as follows. In Sec.~\ref{sec:II}, we derive the expressions of the form factors and decay width in CLFQM. In Sec.~\ref{sec:III}, we show our numerical results and provide a discussion. The conclusion is given in Sec.~\ref{sec:V}.

\section{The decay of $X(3872)\to J/\psi\gamma, \, \psi'\gamma$}
\label{sec:II}
\subsection{Notation}
We take the covariant light-front quark model used in \cite{Jaus:1991cy,Jaus:1989au,Jaus:1996np,Jaus:1999zv,Cheng:2003sm,Choi:2007se,Shi:2016cef,Chang:2019obq} to calculate the width of radiative decay. In Ref.~\cite{Ke:2011jf}, the radiative decay of $X(3872)$ is investigated for the case of $2^{-+}$. In the light-front framework, a momentum $p$ is expressed as
\begin{eqnarray}
p^\mu&=&(p^-, p^+, p_\bot),\nl
p_\bot&=&(p^1, p^2),\,\,  p^-=p^0-p^3,\,\,  p^+=p^0+p^3,
\end{eqnarray}
and thus
\begin{equation}
p^2=p^+p^--p_\bot^2,\,\,\, p\cdot q=\frac{1}{2}(p^-q^++p^+q^-)-p_\bot q_\bot.
\end{equation}
The incoming (outgoing) meson has momentum $P'('') = p_1'('') + p_2$, where $p_1'('')$ and $p_2$ are the momenta of the off-shell active quark and spectator quark, respectively, with masses $m'_1('')$ and $m_2$. We defined $P=P'+P''$ and $q=P'-P''$. An illustration is shown in Fig.~\ref{fig:diagram} \footnote{In principle, there are diagrams of the photon that couples directly to the quark-quark-vector meson vertex. However, as shown in Ref.~\cite{Ganbold:2021nvj} those contributions from the bubble diagrams are small and not more than 10\% of the result. Thus they are omitted without loss of our accuracy for a conclusion}. These momenta can be expressed in terms of the internal variables ($x_i, p'_\bot$),
\begin{eqnarray}
&&p_1'('')^+=x_1 P^{'('')+}, \,\,\, p_2'('')^+=x_2 P'('')^+\nl
&&p'('')_{1\bot}=x_1 P'('')_\bot+p'('')_\bot,\nl
&&p'('')_{2\bot}=x_2 P^{'('')}_\bot-p'('')_\bot,\nl
&&p''_\bot=p'_\bot-x_2 q_\bot,
\end{eqnarray}
with $x_1+x_2=1$.
The following physical quantities are defined for subsequent derivations:
\begin{eqnarray}
&&M_0'^2=(e_1'+e_2)^2=\frac{p_\bot'^2+m_1'^2}{x_1}+\frac{p_\bot'^2+m_2^2}{1-x_1},\nl
&&e_1'=\sqrt{p_\bot'^2+p_z'^2+m_1'^2},\,\,e_2=\sqrt{p_\bot'^2+p_z'^2+m_2^2},\nl
&&p_z'=\frac{(1-x_1)M_0'}{2}-\frac{p_\bot'^2+m_2^2}{2(1-x_1)M_0'},\nl
&&M_0''^2=\frac{p_\bot''^2+m_1''^2}{x_1}+\frac{p_\bot''^2+m_2^2}{1-x_1},\nl
&&e_1''=\sqrt{p_\bot''^2+p_z''^2+m_1''^2},\nl
&&p_z''=\frac{(1-x_1)M_0''}{2}-\frac{p_\bot''^2+m_2^2}{2(1-x_1)M_0''}.
\end{eqnarray}
Here, $M_0'^2$ ($M_0''^2$) can be interpreted as the kinetic invariant mass squared of the incoming (outgoing) $q\overline{q}$ system, and
$e_i$ is the energy of Quark $i$.

\subsection{Form factors}

\begin{figure}[h]
 \centering
 \includegraphics[scale=0.2]{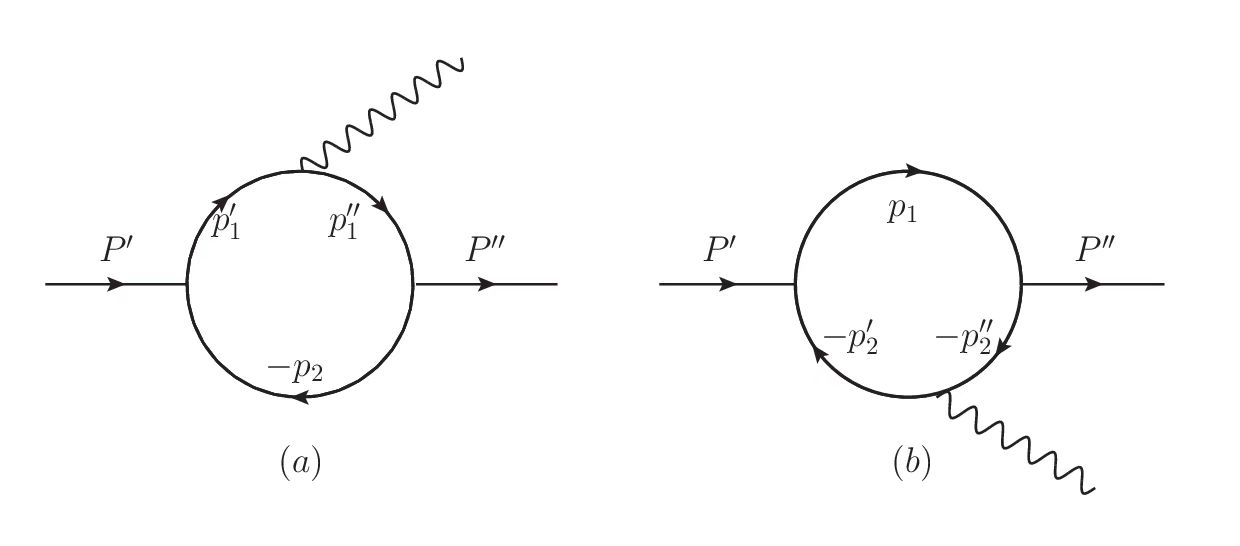}
\caption{Feynman diagrams for radiative transitions. The uppercase $P'$ and $P''$ denote the four-momentum of the initial and final meson, respectively. The lower case $p_{1,2}$ is the momentum of the spectator quark, and other momenta with prime or double prime in the superscript correspond to the active quarks involving a photon emission shown by a wavy line.}\label{fig:diagram}
\end{figure}

The transition amplitude for $X(3872)\to J/\psi\,\gamma, \psi'\gamma$ can be written as
\begin{eqnarray}
&&A(X(3872)\rightarrow \psi\gamma)=\epsilon^{*\alpha}(q)\epsilon'^\mu(P')\epsilon''^{*\nu}(P'')\mathcal{A}_{\alpha\mu\nu}\nl
&&\mathcal{A}_{\alpha\mu\nu}=\varepsilon_{\alpha\nu\beta\eta}P^\beta q^\eta[P_\mu f_1(q^2)+q_\mu f_2(q^2)]\nl
&&\qquad\quad+\varepsilon_{\alpha\mu\beta\eta}P^\beta q^\eta[P_\nu f_3(q^2)+q_\nu f_4(q^2)]\nl
&&\qquad\quad+\varepsilon_{\alpha\mu\nu\rho}[P^\rho f_5(q^2)+q^\rho f_6(q^2)]\nl
&&\qquad\quad+\varepsilon_{\mu\nu\beta\eta}P^\beta q^\eta[P_\alpha f_7(q^2)+q_\alpha f_8(q^2)],
\end{eqnarray}
where $\epsilon'$ denotes the polarization of $X(3872)$, $\epsilon''$ denotes the polarization of $J/\psi$ or $\psi'$, and $\epsilon$ denotes the polarization of the photon. $f_i(q^2)$ with $i=1\cdots 8$ is the $q^2$-dependent form factor that encodes the underlying dynamics. By considering the properties of the polarization vectors, the following can be obtained:
\begin{eqnarray}
&&\epsilon^{*\alpha}(q)\cdot q_\alpha=0\nl
&&\epsilon'^\mu(P')\cdot P'_\mu=\epsilon'^\mu(P')\cdot \frac{1}{2}(P+q)_\mu=0\nl
&&\epsilon''^{*\nu}(P'')\cdot P''_\nu=\epsilon''^{*\nu}(P'')\cdot \frac{1}{2}(P-q)_\nu=0,
\end{eqnarray}
$\mathcal{A}_{\alpha\mu\nu}$ will be reduced to, and only the nonvanishing terms are maintained as follows:
\begin{eqnarray}
\mathcal{A}_{\alpha\mu\nu}&=&\varepsilon_{\alpha\nu\beta\eta}P^\beta q^\eta P_\mu f_m(q^2)+\varepsilon_{\alpha\mu\beta\eta}P^\beta q^\eta P_\nu f_p(q^2)\nl&&+\varepsilon_{\alpha\mu\nu\rho}[P^\rho f_5(q^2)+q^\rho f_6(q^2)]\nl&& +\varepsilon_{\mu\nu\beta\eta}P^\beta q^\eta P_\alpha f_7(q^2),
\end{eqnarray}
where $f_m(q^2)=f_1(q^2)-f_2(q^2)$, $f_p(q^2)=f_3(q^2)+f_4(q^2)$. Imposing the gauge invariance $q^\alpha A_{\alpha\mu\nu}=0$, we further obtained
\begin{eqnarray}\label{form factor}
\mathcal{A}_{\alpha\mu\nu}&=&\varepsilon_{\alpha\nu\beta\eta}P^\beta q^\eta P_\mu f_m(q^2)+\varepsilon_{\alpha\mu\beta\eta}P^\beta q^\eta P_\nu f_p(q^2)\nl&&+\varepsilon_{\alpha\mu\nu\rho}q^\rho f_6(q^2).
\end{eqnarray}

For the calculation of radiative decay, we employed the covariant light-front quark model based on the assumption that $X(3872)$ is regarded as an axial vector ($A$) meson of $c\overline{c}$, while $J/\psi$ ($\psi'$) is a vector ($V$) meson. The vertex functions for the axial vector meson and vector meson states are written as \cite{Cheng:2003sm}
\begin{eqnarray}
&-iH_A'\left(\gamma_\mu+\frac{(p_1'-p_2)_\mu}{W_A}\right)\gamma_5,\nl
&iH_V''\left(\gamma_\mu-\frac{(p_1''-p_2)_\mu}{W_V}\right),
\end{eqnarray}
where $H_A'$ and $H_V''$ are the wave functions of the axial vector meson and vector meson that will be given below. The Feynman amplitude corresponding to Fig.~\ref{fig:diagram} can be written as
\begin{eqnarray}
A_{\alpha\mu\nu}&=&\frac{N_c}{(2\pi)^4}\int d^4p_1'H_A'H_V''\nl &&\times\left(\frac{2e}{3}\frac{S_{\alpha\mu\nu}^a}{N_1'N_1''N_2}+\frac{2e}{3}\frac{S_{\alpha\mu\nu}^b}{{N_2'N_2''N_1}}\right),
\end{eqnarray}
where $N_c=3$, $2e/3$ is the electric charge of the charm quark, $N_i=p_i^2-m_i^2+i0^+$, $N'_i=p'^2_i-m'^2_i+i0^+$, and $N''_i=p''^2_i-m''^2_i+i0^+$. The terms $S_{\alpha\mu\nu}^a$ and $S_{\alpha\mu\nu}^b$ are the traces corresponding to Fig.~1(a) and Fig.~1(b), respectively. They are shown as follows:
\begin{widetext}
\begin{eqnarray}
S_{\alpha\mu\nu}^a&=&\text{Tr}\Big[(\gamma_\mu+\frac{(p_1'-p_2)_\mu}{W_A'})\gamma_5(-p_2\mkern-10.5 mu/+m_2)(\gamma_\nu-\frac{(p_1''-p_2)_\nu}{W_V''})(p_1''\mkern-10.5 mu/+m_1'')\gamma_\alpha(p_1'\mkern-10.5 mu/+m_1')\Big],\nl
S_{\alpha\mu\nu}^b&=&\text{Tr}\Big[(\gamma_\mu+\frac{(p_1-p_2')_\mu}{W_A'})\gamma_5(-p_2'\mkern-10.5 mu/+m_2')\gamma_\alpha(-p_2''\mkern-10.5 mu/+m_2'')(\gamma_\nu-\frac{(p_1-p_2'')_\nu}{W_V''})(p_1\mkern-10.5 mu/+m_1)\Big].
\end{eqnarray}
Explicitly, the following can be obtained:
\begin{eqnarray}
S_{\alpha\mu\nu}^a&=&\varepsilon_{\alpha\mu\nu\rho}\Big[-2p_1'^\rho(N_1''-2m_2m_1''+m_1''^2-2m_1''m_1'+N_1'-q^2)+q^\rho(N_1''+4m_2m_1'\nl
&&+m_1''^2-2m_1''m_1'+m_1'^2+N_1'-q^2)+P^\rho(N_1''+m_1''^2-2m_1''m_1'+m_1'^2+N_1'-q^2)\Big]\nl
&&+2\varepsilon_{\alpha\mu\beta\eta}q^\beta p_1'^\eta\Big(P_\nu+q_\nu-2p_{1\nu}'\Big)+2\varepsilon_{\alpha\nu\beta\eta}q^\beta p_1'^\eta\Big(P_\mu+q_\mu-2p_{1\mu}'\Big)\nl
&&+\varepsilon_{\mu\nu\beta\eta}\Big[2P^\beta p_1'^\eta(2p_{1\alpha}'-q^\alpha)-2P^\beta q^\eta p_{1\alpha}'-2q^\beta p_1'^\eta q_\alpha\Big]\nl
&&+\frac{\varepsilon_{\alpha\nu\beta\eta}}{W_A'}(P_\mu+q_\mu-4p_{1\mu}')
\Big[p_1'^\eta q^\beta(m_1'+m_1''-2m_2)+P^\beta p_1'^\eta(m_1''+m_1')+P^\beta q^\eta m_1'\Big]\nl
&&+\frac{\varepsilon_{\alpha\mu\beta\eta}}{W_V''}(P_\nu+3q_\nu-4p_{1\nu}')
\Big[-p_1'^\eta q^\beta(m_1'+m_1''+2m_2)+P^\beta p_1'^\eta(+m_1'-m_1'')-P^\beta q^\eta m_1'\Big]\nl
&&+\frac{\varepsilon_{\alpha\beta\eta\rho }P^\beta q^\eta p_1'^\rho(P_\nu+3q_\nu-4p_{1\nu}')(P_\mu+q_\mu-4p_{1\mu}')}{2 W_A' W_V''}.
\end{eqnarray}
\end{widetext}
In the light-front formalism, the integration measure is
\begin{equation}\label{zhuanhuan}
d^4p'_1=\frac{1}{2}P'^+dp_1'^-dx_1 d^2p'_\bot.
\end{equation}
As discussed in Ref.~\cite{Jaus:1999zv}, if it is assumed that the vertex function ($H_A'$, $H_V''$) has no pole in the upper complex $p_1'^-$ plane, the CLFQM used here and the standard light-front formulism where the covariance property is not satisfied will lead to identical results at the one-loop level. We assume such a situation holds here. Then, the integration over $dp_1'^-$ picks up a residue at $p_2=\hat{p}_2$, where $\hat{p}_2^2=m_2^2$. Thus, the following replacements will be obtained:
\begin{eqnarray}\label{1}
&&N_1'\rightarrow \hat{N}_1'=x_1(M'^2-M_0'^2),\nl
&&N_1''\rightarrow \hat{N}_1''=x_1(M''^2-M_0''^2),\nonumber\\
&&H_A'\rightarrow \hat{H}_A'\equiv h_A', \quad H_V''\rightarrow \hat{H}_V''\equiv h_V'',\nonumber\\
&&W_A'\rightarrow \hat{W}_A'\equiv w_A', \quad W_V''\rightarrow \hat{W}_V''\equiv w_V'',\nonumber\\
&&\int d^4p_1'\frac{H_A'H_V''}{N_1'N_1''N_2}S_{\alpha\mu\nu}^a \nl
&&\qquad \rightarrow -i\pi\int dx_2d^2p_\perp'\frac{h_A'h_V''}{x_2\hat{N}_1'\hat{N}_1''}\hat{S}_{\alpha\mu\nu}^a,
\end{eqnarray}
with \cite{Cheng:2003sm}
\begin{eqnarray}
&&h_A'=\sqrt{\frac{3}{2}}(M'^2-M_0'^2)\sqrt{\frac{x_1x_2}{N_c}}\frac{1}{\sqrt{2}\widetilde{M}_0'}\frac{\widetilde{M}_0'^2}{2\sqrt{3}M_0'}\varphi'(nP),\nl
&&h_V''=(M''^2-M_0''^2)\sqrt{\frac{x_1x_2}{N_c}}\frac{1}{\sqrt{2}\widetilde{M}_0''}\varphi''(nS),\nl
&&w_A' = \frac{\widetilde{M}_0'^2}{m_1'-m_2},  \quad w_V''= M_0''+m_1''+m_2,\nl
&&\widetilde{M}_0'('') = \sqrt{M_0'('')^2-(m_1'('')-m_2)^2}.
\end{eqnarray}
Considering $X(3872)$ as a $2P$ charmonium, we need the following wave functions ($1S$ for $J/\psi$ and $2S$ for $\psi'$) \cite{Hwang:2008qi}
\begin{eqnarray}\label{eq:wf}
\varphi(1S)&=&4\left(\frac{\pi}{\beta^2}\right)^{\frac{3}{4}}\sqrt{\frac{dp_z}{dx_2}}\exp\left(-\frac{p_z^2+p_\bot^2}{2\beta^2}\right),\nl
\varphi(2S)&=&4\sqrt{\frac{8}{3}}\left(\frac{\pi}{\beta^2}\right)^{\frac{3}{4}}\sqrt{\frac{dp_z}{dx_2}}\left(\frac{p_z^2+p_\bot^2}{2\beta^2}-\frac{3}{4}\right)\nl
&&\times\exp\left(-\frac{p_z^2+p_\bot^2}{2\beta^2}\right),\nl
\varphi(2P)&=&4\left(\frac{\pi}{\beta^2}\right)^{\frac{3}{4}}\sqrt{\frac{dp_z}{dx_2}}\sqrt{\frac{5}{\beta^2}}\left(1-\frac{2}{5}\frac{p_z^2+p_\bot^2}{2\beta^2}\right)\nl
&&\times\exp\left(-\frac{p_z^2+p_\bot^2}{2\beta^2}\right). 
\end{eqnarray}
with
\begin{equation}
\frac{dp_z'('')}{dx_2}=\frac{e_1'('')e_2}{x_1x_2M_0'('')}.
\end{equation}
For the practical calculation, the symbols $p_z$ and $p_\bot$ in Eq.~\eqref{eq:wf} should be associated with the superscript $'$ for the initial meson and $''$ for the final meson.

In a concrete calculation in the light front quark model, one usually works in the $q^+=0$ frame, where one only needs to consider the
valence quark part, and all other non-valence contributions e.g., the pair creations from the vacuum (the so-called Z diagram) vanish.
By making the contour integral over $p_1'^-$ for the four-dimensional integration, one gets the above hat quantities. These matrix elements will acquire a spurious $\omega$ dependence, with $\omega=(2, 0, 0_\perp)$ as a constant. Thus the covariance is lost. The reason is due to the missing contribution in the $p_1'^-=p_1''^-=0$ region. Once such contribution is correctly included, the covariance will be restored. W.~Jaus finds an effective way for inclusion of such zero mode contribution as the necessary condition for the covariance, and the net results are \cite{Jaus:1999zv}:
\begin{eqnarray}
\hat{p}'_{1\mu}\rightarrow&& P_\mu A_1^{(1)}+q_\mu A_2^{(1)}\nl
\hat{p}'_{1\mu}\hat{p}'_{1\nu}\rightarrow&& g_{\mu\nu}A_1^{(2)}+P_\mu P_\nu A_2^{(2)}+(P_\mu q_\nu+q_\mu P_\nu) A_3^{(2)}\nl
&&+q_\mu q_\nu A_4^{(2)}\nl
\hat{p}'_{1\mu}\hat{p}'_{1\nu}\hat{p}'_{1\alpha}\rightarrow&& (g_{\mu\nu}P_\alpha+g_{\mu\alpha}P_\nu+g_{\nu\alpha}P_\mu)A_1^{(3)}\nl
&&+(g_{\mu\nu}q_\alpha+g_{\mu\alpha}q_\nu+g_{\nu\alpha}q_\mu)A_2^{(3)}\nl
&&+P_\mu P_\nu P_\alpha A_3^{(3)}\nl
&&+(P_\mu P_\nu q_\alpha+P_\mu q_\nu P_\alpha+q_\mu P_\nu P_\alpha)A_4^{(3)}\nl
&&+(q_\mu q_\nu P_\alpha+q_\mu P_\nu q_\alpha+P_\mu q_\nu q_\alpha) A_5^{(3)} \nl
&&+q_\mu q_\nu q_\alpha A_6^{(3)},
\end{eqnarray}
with those coefficients $A_i^{(j)}$ given by
\begin{equation}
\begin{split}
&A_1^{(1)}=\frac{x_1}{2}, \,\, A_2^{(1)}=\frac{x_1}{2}-\frac{p'_\bot \cdot q_\bot}{q^2}\\
&A_1^{(2)}=-p'^2_\bot-\frac{(p'_\bot \cdot q_\bot)^2}{q^2}, \,\,  A_2^{(2)}=(A_1^{(1)})^2\\
&A_3^{(2)}=A_1^{(1)}A_2^{(1)}, \,\, A_4^{(2)}=(A_2^{(1)})^2-\frac{A_1^{(2)}}{q^2}\\
&A_1^{(3)}=A_1^{(1)}A_1^{(2)}, \,\, A_2^{(3)}=A_1^{(2)}A_2^{(1)}\\
&A_3^{(3)}=A_1^{(1)}A_2^{(2)}, \,\, A_4^{(3)}=A_2^{(1)}A_2^{(2)}\\
&A_5^{(3)}=A_1^{(1)}A_4^{(2)}, \,\, A_6^{(3)}=A_2^{(1)}A_4^{(2)}-\frac{2}{q^2}A_2^{(1)}A_1^{(2)}\\
\end{split}
\end{equation}

By matching with Eq.~\eqref{form factor}, the expressions of the form factors can be obtained as follows:
\begin{widetext}
\begin{eqnarray}\label{eq:ff}
f_m^a(q^2)&=&\frac{2e}{3}\frac{N_c}{16\pi^3}\int dx_2d^2p_\perp'\frac{h_A'h_V''}{x_2\hat{N}_1'\hat{N}_1''}(-4)\left\{\frac{1}{w_A'}\Big[(m_1'-m_1'')(A_3^{(2)}-A_4^{(2)})\right.\nl
&&+(m_1'+m_1''-2m_2)\times(A_2^{(2)}-A_3^{(2)})+m_1'(A_2^{(1)}-A_1^{(1)})\Big]+A_2^{(2)}-A_3^{(2)}\nl
&&\left.-\frac{1}{w_A'w_V''}(2A_2^{(3)}-2A_1^{(3)})\right\},\nl
f_p^a(q^2)&=&\frac{2e}{3}\frac{N_c}{16\pi^3}\int dx_2d^2p_\perp'\frac{h_A'h_V''}{x_2\hat{N}_1'\hat{N}_1''}(-4)\left\{\frac{1}{w_V''}\Big[(m_1'-m_1'')(A_3^{(2)}+A_4^{(2)}\right.\nl
&&-A_2^{(1)})+(m_1'+m_1''+2m_2)\times(A_2^{(2)}+A_3^{(2)}-A_1^{(1)})-m_1'(A_1^{(1)}+A_2^{(1)}\nl
&&\left.-1)\Big]+A_1^{(1)}-A_2^{(2)}-A_3^{(2)}-\frac{1}{w_A'w_V''}(2A_1^{(3)}+2A_2^{(3)}-2A_1^{(2)})\right\},\nl
f_6^a(q^2)&=&\frac{2e}{3}\frac{N_c}{16\pi^3}\int dx_2d^2p_\perp'\frac{h_A'h_V''}{x_2\hat{N}_1'\hat{N}_1''}(-4)\left\{\frac{1}{w_A'}(m_1'+m_1''-2m_2)A_1^{(2)}\right.\nl
&&+\frac{1}{w_V''}(m_1'+m_1''+2m_2)A_1^{(2)}-\frac{1}{4}(1-2A_2^{(1)})\Big[-q^2+\hat{N}_1'+\hat{N}_1''\nl
&&\left.+(m_1'-m_1'')^2\Big]-A_2^{(1)}(m_1''m_2-m_1'm_2)-m_1'm_2\hspace{-4ex}\phantom{\frac{1}{w_A'}}\right\}.
\end{eqnarray}
\end{widetext}
The form factors corresponding to Fig.~1(b), i.e., the expressions of $f_m^b(q^2),f_p^b(q^2),f_6^b(q^2)$, can be obtained from Eq.~\eqref{eq:ff} by the interchanges
\begin{equation}
m_1'\rightarrow m_2',\, m_1''\rightarrow m_2'',\, m_2\rightarrow m_1.
\end{equation}
In practice, all these quark masses are the mass of the charm quark, and the contribution of Fig.~1(b) is the same as that of Fig.~1(a).
The form factors will finally be
\begin{eqnarray}
f_i(q^2)=f_i^a(q^2)+f_i^b(q^2)=2f_i^a(q^2),
\end{eqnarray}
with $i$ denoting the indices $m,\,p$, and 6. For the radiative decay considered here, only the form factor values at $q^2=0$ are relevant.

\subsection{Decay width}
The decay width will be conveniently expressed in the helicity basis. We then define the helicity amplitude as $A_{\lambda'\lambda''\lambda_\gamma}$,
with $\lambda', \, \lambda'', \, \lambda_\gamma$ denoting the helicity of $X(3872)$, $J/\psi$ (or $\psi'$), and the photon, respectively. In the rest frame of the initial meson $X(3872)$, we can obtain the explicit representations of the momenta and polarization vectors as follows \cite{Dubnicka:2011mm}:
\begin{eqnarray}
&&P'^\mu=(M',0,0,0), \quad P''^\mu=(E_1,0,0,|\bm q|), \nl
&&q^\mu=(|\bm q|,0,0,-|\bm q|),\nl
&&\epsilon'^{\mu}_\pm=\frac{1}{\sqrt{2}}(0,\mp1,-i,0),\,\, \epsilon'^{\mu}_0=(0,0,0,1),\nl
&&\epsilon''^{\mu}_\pm=\frac{1}{\sqrt{2}}(0,\mp1,-i,0),\,\, \epsilon''^{\mu}_0=\frac{1}{M''}(|\bm q|,0,0,E_1),\nl
&&\epsilon^{\mu}_\pm(\gamma)=\frac{1}{\sqrt{2}}(0,\pm1,-i,0),
\end{eqnarray}
where $|{\bm q}|=\frac{M'^2-M''^2}{2M'}$ and $E_1=\frac{M'^2+M''^2}{2M'}$ is the energy of the vector meson.
Due to the conservation of angular momentum, $\lambda'=\lambda''-\lambda_\gamma$, and the nonvanishing amplitudes for $X(3872)\to J/\psi \gamma,\,\psi'\gamma$ decay will be
\begin{eqnarray}\label{eq:E1M2}
A_{+0-}&=&-i\frac{M'|\bm q|}{M''}(f_6+2M'|\bm q|f_p),\nl
A_{0++}&=&i |\bm q| (f_6+2 M' |\bm q| f_m),\nl
A_{-0+}&=&-A_{+0-},\quad A_{0--}=-A_{0++}.
\end{eqnarray}
The last two equations follow from the parity relation and have also been verified by calculation.
Obviously, we can also adopt the convention of the polarization vectors used in Refs.~\cite{Zhang:2020dla,Ivanov:2019nqd}. The difference lies only in the definition of the momentum direction and the resulting change in the polarization vector forms. We verified that they give identical physical results. Stated differently, this transition includes the $E1$ and $M2$ types, which are characterized by the $\bm q$ and $\bm q^2$ for the near-threshold behavior for the amplitude. Of course, both $S$ and $D$ waves are included, and combined together in Eq.~\eqref{eq:E1M2}. The helicity relation $\lambda'=\lambda''-\lambda_\gamma$ does not imply $S$ wave solely, since the projection of any orbital angular momentum onto the linear momentum vanishes.

The decay width can be calculated by
\begin{eqnarray}\label{decay width}
\Gamma&=&\frac{|\bm q|}{8\pi M'^2}\left(|A_{+0-}|^2+|A_{-0+}|^2+|A_{0--}|^2+|A_{0++}|^2\right)\nl
&=&\frac{|\bm q|^3}{\pi}\left(\frac{f_6+2M'|\bm q|f_p}{4M''^2}+\frac{f_6^2}{4M'^2}+\frac{f_6f_m|\bm q|}{M'}+f_m^2|\bm q|^2\right).\nl
\end{eqnarray}
The dimension of the amplitude $A_{\lambda'\lambda''\lambda_\gamma}$ is [mass]$^1$, while the form factors $f_6, f_p$, and $f_m$ have dimensions of [mass]$^0$, [mass]$^{-2}$ and [mass]$^{-2}$, respectively.

\section{numerical results and discussions}
\label{sec:III}
By using Eqs.~\eqref{eq:ff} and \eqref{decay width} in Sec.~\ref{sec:II}, we will calculate the width of radiative decay. We take the constituent quark mass $m_c=1.4$ GeV \cite{Cheng:2003sm,Jaus:1989au,Jaus:1996np} for calculation. In the wave function of a meson, there is a parameter $\beta$ that needs to be determined. For vector mesons (V) $J/\psi$ and $\psi'$, we fix the $\beta$ value by their decay constants, which are extracted from the decay width to the $e^+e^-$ pair through the following equation:
\begin{equation}
\Gamma(V\rightarrow e^+e^-)=\frac{4\pi}{3}\frac{4}{9}\alpha^2 \frac{f_V^2}{M'},
\end{equation}
where $\alpha$ is the fine structure constant, $f_V$ is the decay constant, and $M'$ is the mass of the vector meson. By taking $\text{Br}(J/\psi\rightarrow e^+e^-)=(5.971\pm0.032)\%$, $\text{Br}(\psi'\rightarrow e^+e^-)=(7.93\pm0.17)\times10^{-3}$, $\Gamma(J/\psi)=(92.6\pm1.7)$ keV and $\Gamma(\psi(2S))=(294\pm8)$ keV \cite{Workman:2022ynf}, we can obtain the decay constants $f_{J/\psi}=415.49$ MeV and $f_{\psi'}=294.35$ MeV as our central values. The uncertainties are very small. The formula for the decay constant of vector mesons in the LFQM is given by \cite{Cheng:2003sm}
\begin{eqnarray}
f_V&=&\frac{N_c}{4\pi^3M'}\int dx d^2p'_\bot\frac{h_V'}{x(1-x)(M'^2-M_0'^2)}\nl&&\times\Big[xM_0'^2 -m_1'(m_1'-m_2)-p_\bot'^2\nl
&&\quad+\frac{m_1'+m_2}{M_0'+m_1'+m_2}p_\bot'^2\Big],
\end{eqnarray}
from which we fix the parameters $\beta_{J/\psi}=0.631$ GeV and $\beta_{\psi'}=0.487$ GeV.

Nowadays the PDG reported the $X(3872)$ width as $\Gamma(X(3872)=(1.19\pm0.21)$ MeV \cite{Workman:2022ynf} based on the LHCb measurements \cite{LHCb:2020fvo,LHCb:2020xds}. But this corresponds to the Breit-Weigner (BW) width. The BW parametrization is not an appropriate parametrization for a very near-threshold state. LHCb collaboration has discussed their pole searches \cite{LHCb:2020xds} and very recently BESIII did such activity too \cite{BESIII:2023hml}. We will take the central value of 0.4 MeV for the X(3872) pole width \cite{BESIII:2023hml} for illustration despite the large uncertainties in the pole parameters, and this width value also agrees with the reported full width at the half maximum for the line shape \cite{BESIII:2023hml}. Taking the branching ratio values $\text{Br} (X(3872)\rightarrow \psi'\gamma)=(4.5\pm2.0)\%$ and $\text{Br} (X(3872)\rightarrow J/\psi\gamma)=(8\pm4)\times 10^{-3}$ \cite{Workman:2022ynf}, we obtain $\Gamma_{\text{exp}}(X(3872)\rightarrow \psi'\gamma)=(1.8\pm0.8) \times 10^{-2}$ MeV and $\Gamma_{\text{exp}}(X(3872)\to J/\psi \gamma)=(3.2\pm1.6)\times 10^{-3}$ MeV as the ``experimental'' values for the partial decay widths. Those uncertainties purely come from the ones for the branching ratios. From the partial decay width to $\psi'\gamma$, we are able to fix the $\beta$ value for $X(3872)$, $\beta_{X(3872)}$, as $0.56^{+0.04}_{-0.03}$ GeV. Then the theoretical value for $\Gamma(X(3872)\to J/\psi\gamma)$ will be predicted to $(9.1^{+1.7}_{-1.5})\times10^{-1}$ MeV, which deviates the aforementioned ``experimental'' number too much. Consequently, the scenario of a pure $c\overline c$ assignment for $X(3872)$ will encounter difficulty in reconciling the widths to $J/\psi\gamma$ and $\psi'\gamma$.

Here we clarify more on the uncertainty for our result. The central value of $\beta_{X(3872)}$ is required to reproduce the central value of the $\Gamma(X(3872)\rightarrow \psi' \gamma)$, while its asymmetric errors are obtained by sampling $\Gamma(X(3872)\rightarrow \psi' \gamma)$ within one standard deviation range. From the produced set of $\beta_{X(3872)}$ numbers, the maximum and minimum values are picked out. Obviously, the value of $\Gamma(X(3872)\to J\psi \gamma)$ takes into account the uncertainty of $\beta_{X(3872)}$ through the set of the numbers for $\beta_{X(3872)}$. Moreover, we also associate $\Gamma(X(3872)\to J\psi \gamma)$ with another uncertainty by a roughly 10\% of its central value due to the neglected contribution from the $\gamma$-quark-antiquark-meson vertex (see the footnote).

The above results show that assigning $X(3872)$ by a $c\bar c$ state is much disfavored from the viewpoint of its radiative decays. However, other configurations are possible. For example, there is indeed a calculation of radiative decay based on the tetraquark assumption in Ref.~\cite{Dubnicka:2011mm}. There they find a consistency of their theoretical prediction of $\Gamma(X(3872)\to J/\psi \gamma)/\Gamma(X(3872)\to J/\psi 2\pi)$ with the available experimental measurements, by choosing a reasonable value of the size parameter of the $X(3872)$ meson (as a parameter of their model). But note that the calculated quantity therein is the decay width ratio of $\Gamma(X(3872)\to J/\psi \gamma)/\Gamma(X(3872)\to J/\psi 2\pi)$, and here we concern the ratio of $\Gamma(X(3872)\to J/\psi \gamma)/(X(3872)\to \psi' \gamma)$.
The calculation from the $D\bar D^*$ molecular assumption was done in Refs.~\cite{Aceti:2012cb,Guo:2014taa}. The calculation from the coupled channels $D\bar D^* + c \bar c$ was done in Ref.~\cite{Badalian:2012jz}. Future works along the line of investigating the nature of $X(3872)$ through its radiative decays are still meaningful and encouraged.

\section{Conclusion}
\label{sec:V}
The experimental measurements on the $X(3872)$ decay width and branching ratios have made great progress. Assuming that $X(3872)$ is regarded as a conventional charmonium with the quantum number $J^{PC}=1^{++}$, we calculated the transition form factors for $X(3872)$ decaying to a photon and $J/\psi$ ($\psi'$) in the framework of the covariant light-front quark model. In this approach, we need to determine the values of $\beta$ appearing in the wave functions of mesons, where $\beta_{J/\psi}$ and $\beta_{\psi'}$ are fixed by their decay constants, extracted from their partial decay widths to the $e^+e^-$ pair. $\beta_{X(3872)}$ will be fixed by the width of $\Gamma(X(3872)\rightarrow \psi'\gamma)$ reported by the PDG. In this manner, we can cleanly predict the decay width $\Gamma(X(3872)\rightarrow J/\psi\gamma)$. It turns out that the predicted partial decay width of $\Gamma(X(3872)\rightarrow J/\psi\gamma)$ is too large to assign $X(3872)$ as a traditional charmonium state. Or stated differently, the probability that $X(3872)$ is a pure $c\overline{c}$ resonance is rather small.

\section*{Acknowledgment}
We thank Prof.~Yu-Bing Dong, Prof.~Xiang Liu for discussions, and Prof.~V. O. Galkin for his careful reading and suggestions. Prof.~Hong-Wei Ke should be especially acknowledged for his patient discussions in the early stage of our work.
This work is supported by the National Natural Science Foundation of China (NSFC) under Project No. 12275023.

\bibliographystyle{unsrt}
\bibliography{1}

\end{document}